# Synergistic Antenna-Modulator Integration for Monolithic Photonic RF Receiver

Changlin Liu[1,2,†], Yongtao Du[1,2,†], Xihua Zou[1,2,*], Fang Zou[3], Jiejun Zhang[4], Junkai Zhang[4], Xiaojun Xie[1,2], Wei Pan[1,2], Lianshan Yan[1,2], and Jianping Yao[5]

[1]Center for Information Photonics and Communications, School of Information Science and Technology, Southwest Jiaotong University, Chengdu, China

[2]Key Laboratory of Photonic-Electric Integration and Communication-Sensing Convergence (Ministry of Education), Chengdu, China

[3]Tianfu Xinglong Lake Laboratory, Chengdu, China

[4]College of Physics & Optoelectronic Engineering, Jinan University, Guangzhou, China

[5]School of Electrical Engineering and Computer Science, University of Ottawa, Ottawa, Canada

[†]These authors contributed equally: Changlin Liu, Yongtao Du

*Corresponding author: zouxihua@swjtu.edu.cn

**Abstract:** Integrated radio-frequency (RF) photonics plays a pivotal role in wireless communications, sensing, and radar due to its large intrinsic bandwidth, remote distribution capability, and compact footprint. However, despite significant advances in photonic integrated circuits (PICs), the practical deployment of these systems remains constrained by the bulky nature of essential RF components (e.g., antennas, amplifiers, and cables), especially in covert, conformal, and space-constrained applications. To overcome these limitations, monolithic electronic-photonic integrated circuits (EPICs), enabling miniaturized and synergistic integration of both RF and photonic components, are gaining notable attention. As a groundbreaking advancement, we demonstrate a novel photonic RF receiver that monolithically integrates a bow-tie antenna and a microring modulator on a thin-film lithium niobate platform. The chip innovatively leverages dual-resonance enhancement mechanism, RF resonance from the antenna and optical resonance from the microring, to significantly boost the RF-to-optical conversion efficiency. A record-high figure of merit (FOM) of 3.88 $W^{-1/2}$ is achieved within a compact footprint of 2×1.7 $mm^2$. As the first demonstrations, the integrated receiver is deployed in an integrated sensing and communication (ISAC) system, achieving centimeter-level radar ranging accuracy and Gbps-level wireless communication capacity, as well as real-time video transmission function in moving scenarios. This seminal work paves a new way for covert, conformal, and miniaturized frontends in wireless communication and sensing applications, including body area networks, unmanned aerial vehicles, high-speed vacuum maglevs, and electronic warfare systems.

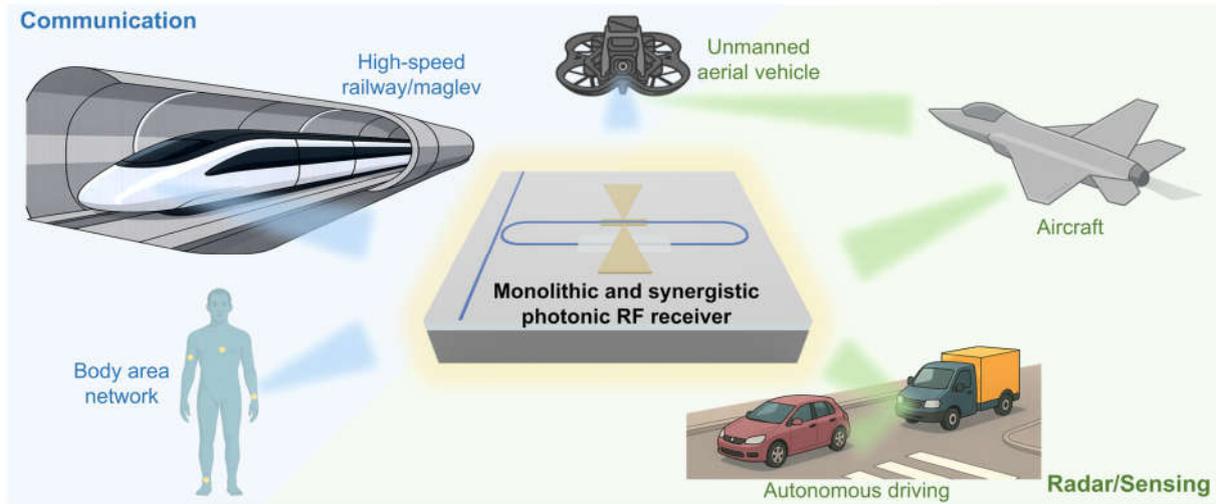

**Fig. 1. Prospective applications of the monolithic and synergistic photonic RF receiver in wireless communication, sensing, and radar scenarios.**

## 1. Design principle

To make RF receivers suitable for wireless communication and radar/sensing scenarios, including body area networks, autonomous vehicles, high-speed trains, vacuum maglevs, and electronic warfare, we propose a monolithic photonic RF receiver with synergistic integration of RF antenna and electro-optic modulator (Fig. 1). On the chip, a bow-tie antenna with two lumped electrodes is co-designed and co-integrated with a high-Q optical microring modulator. A dual-resonance enhancement mechanism is proposed here, which merges the concentrated electromagnetic field from the antenna's inductor-capacitor (LC) resonance and the prolonged photon lifetime from the microring resonance. Thanks to this dual-resonance enhancement mechanism, the overall RF-to-optical conversion efficiency is dramatically enhanced within a compact footprint.

As the first part of the dual-resonance enhancement mechanism, the bow-tie antenna can be approximately modelled as an LC resonator whose resonant frequency is determined by geometric parameters. Although longer electrodes can induce larger refractive index change and higher electro-optic modulation efficiency, the maximum feasible electrode length is usually constrained by the desired resistor-capacitor (RC) time constant for sustaining a high operation frequency[1]. Based on this criterion, the bow-tie antenna and electrodes are co-optimized to achieve a resonant frequency of 30.5 GHz.

As the second part of the dual-resonance enhancement mechanism, the microring cavity introduces optical resonance that is synergetic with the antenna resonance. The photon lifetime in a high-Q microring resonator is highly prolonged, enabling the optical carrier to circulate numerous times within the cavity. By carefully matching the free spectral range (FSR) to the RF frequency, optical sidebands generated from repeated electro-optic modulation and successive round trips constructively interfere, leading to cumulative enhancement in modulation depth. A racetrack-shaped microring resonator is precisely configured to fix the FSR at 30.5 GHz that exactly matches the target RF frequency.

## 2. Fabrication and characterization

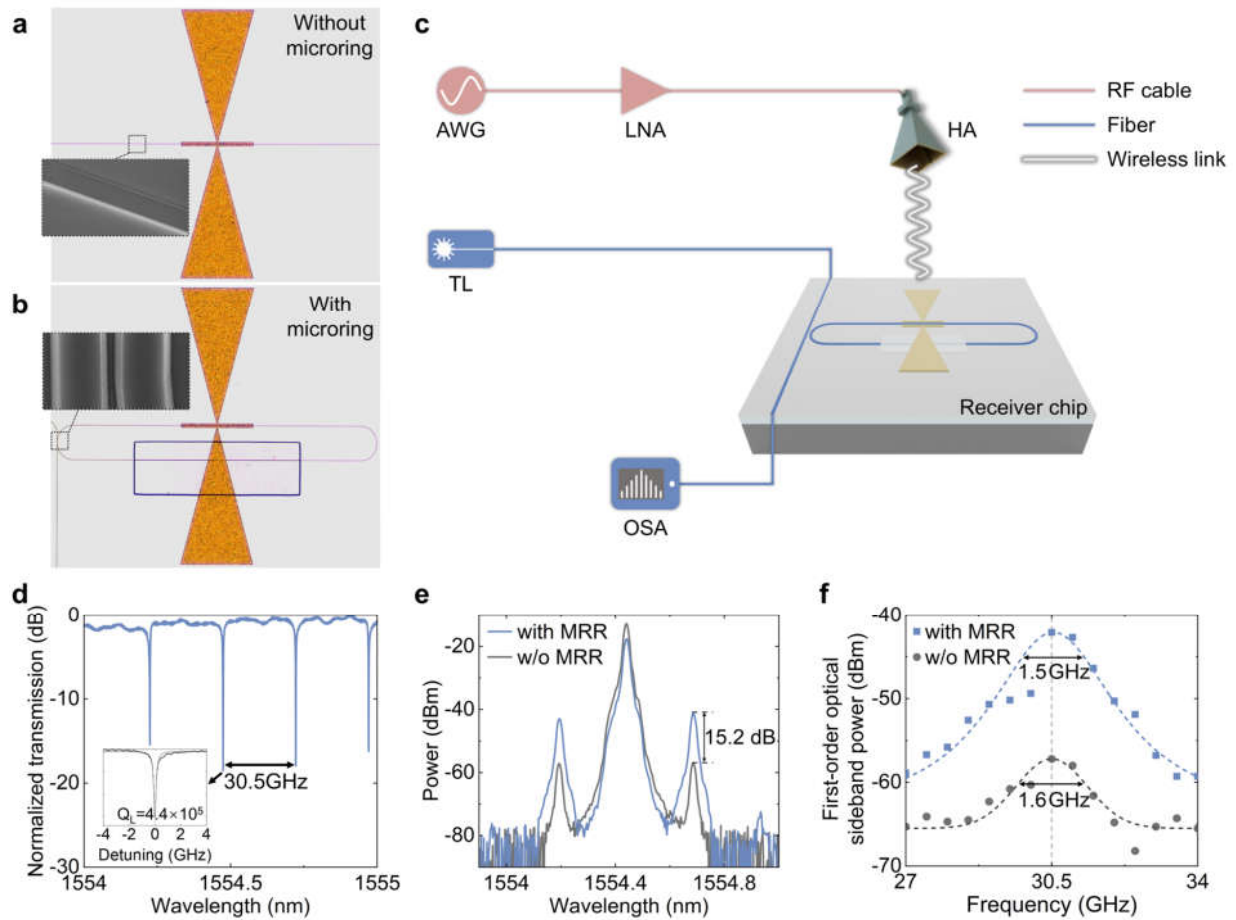

**Fig. 2. Measurement of the monolithic and synergistic receiver chip. a**, **b** Micrographs of the fabricated monolithic and synergistic receiver chips, for (**a**) without and (**b**) with the microring resonator. The insets show the SEM images of the straight waveguide and the coupling region of the microring cavity. **c** Test setup for chip characterization. **d** Measured transmission spectrum of the microring resonator. **e** Optical spectra measured for two configurations with and without the microring under the illumination of a 30.5-GHz RF signal. **f** First-order optical sideband powers obtained at different RF frequencies in both configurations. AWG, arbitrary waveform generator; LNA, low noise amplifier; HA, horn antenna; TL, tunable laser; OSA, optical spectrum analyzer; MRR, microring resonator.

The fabricated chip is shown in Fig. 2a,b, wherein the insets provide scanning electron microscope (SEM) images of the straight waveguide and the coupling region of the microring. The chip is fully characterized using the test setup depicted in Fig. 2c. Firstly, the optical transmission spectrum of the microring is measured through optical wavelength scanning, and the measured FSR successfully aligns the target frequency of 30.5 GHz (Fig. 2d). Next, a single-tone RF signal is generated with its frequency scanning from 27 to 34 GHz, amplified and then radiated into free space through a horn antenna. The received power density (S) incident on the chip is estimated to be 3.5 W·m$^{-2}$. To reveal the significant role of the microring cavity, we compare the optical spectra of two configurations with and without the microring (Fig. 2e). Benefiting from the enormous rounds of electro-optic modulation within the cavity, the power of the first-order optical sidebands increases by 15.2-dB with a slight reduction in the 3-dB bandwidth (Fig. 2f).

The performance of the monolithic and synergistic chip can be fully evaluated from multiple parameter dimensions. A more comprehensive metric for the quantitative evaluation of RF-to-optical conversion efficiency is the FOM[2]. In this case, the proposed monolithic chip strikes a record-high FOM of 3.88 W$^{-1/2}$. Moreover, the synergistic integration of a bow-tie antenna and a microring modulator can be easily extended for large-scale array integration, due to its low loss and compact size.

## 3. Application demonstrations

Benefiting from the outstanding comprehensive specifications above, the monolithic photonic RF receiver is deployed in two promising scenarios for the first time—6G networks and high-speed transportation systems. In the integrated sensing and communication (ISAC) scenario of 6G networks, the synergistic chip simultaneously serves as the receiver for radar sensing and wireless communication. In fast time-varying scenarios like high-speed trains or vacuum maglevs, the vehicle-borne receiver chip suitable for small installation room consistently sustains stable real-time HD video streaming.

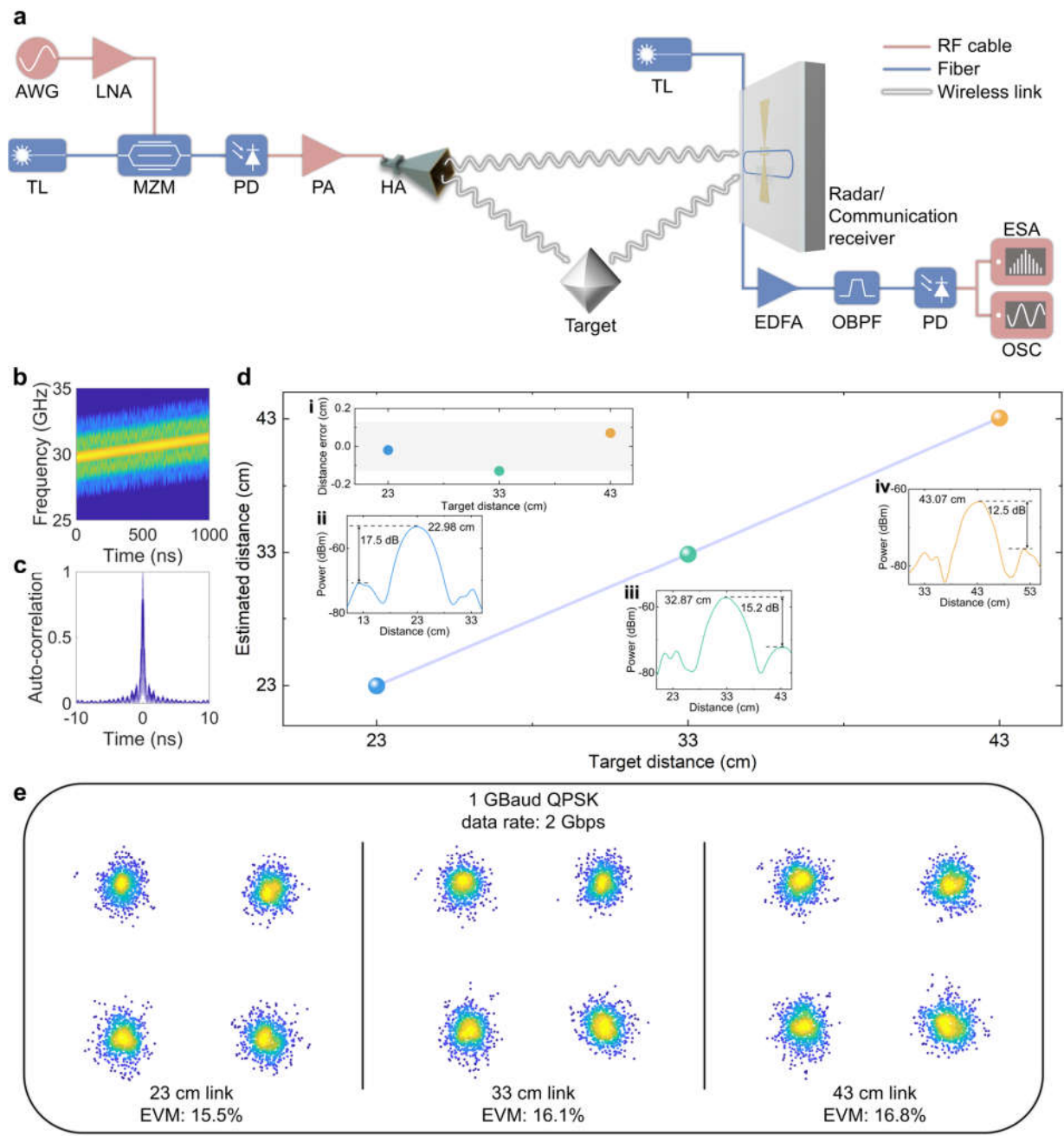

**Fig. 3. Demonstration of a photonic RF ISAC system. a** Experimental setup of the RF ISAC demonstration. **b**, **c** Time-frequency spectrogram (**b**) and auto-correlation characteristic (**c**) of the LFM-QPSK RF signal generated by AWG. **d** Measured distance when the target is placed at 23 cm, 33

cm, and 43 cm away from the antenna. Insets: i) measured distance errors, ii-iv) de-chirped electrical spectra under different target distances. **e** Recovered constellation diagrams of communication signals under different distances. MZM, Mach-Zehnder modulator; PD, photodetector; PA, power amplifier; EDFA, erbium-doped fiber amplifier; OBPF, optical bandpass filter; ESA, electrical spectrum analyzer; OSC, oscilloscope.

**RF integrated sensing and communication system.** Given the growing demand for high-speed communication and high-precision environmental sensing, ISAC has emerged as a new key enabling technology for 6G networks to provide dual functions[3,4]. Here, an RF ISAC system employing the monolithic chip as receiver for both radar and communication is illustrated in Fig. 3a. A continuous-wave optical carrier from a tunable laser is modulated by the ISAC signal inside a Mach-Zehnder modulator. The ISAC waveform comprises a linear frequency modulation (LFM) RF carrier that is phase modulated by a quadrature phase shift keying (QPSK) signal, yielding a constant-envelope LFM–QPSK signal[5]. The corresponding time-frequency spectrogram and normalized auto-correlation characteristic of the LFM–QPSK signal are shown in Fig. 3b,c. After photodetection, the modulated optical signal is converted back to the RF signal and radiated out through a horn antenna, performing radar measurement function using the LFM component and wireless communication function using the QPSK component.

In radar function mode, the RF echo reflected from the target is captured by the chip to modulate the optical carrier. After the de-chirping in optical domain, an intermediate-frequency (IF) signal is generated to estimate the target range. The distance error are 0.02 cm, 0.13 cm, and 0.07 cm, respectively (Fig. 3d).

In wireless communication function mode, the RF signal is also captured by the on-chip bow-tie antenna and applied to modulate the optical carrier within the microring modulator. After optical-to-electrical conversion in a photodetector, the signal is then sampled by a high-speed oscilloscope and demodulated off-line. The constellation diagrams of the recovered QPSK signal are shown in Fig. 3e, and the calculated error vector magnitudes (EVMs) remain below the 17.5% threshold of QPSK signal defined by the 3rd Generation Partnership Project (3GPP).

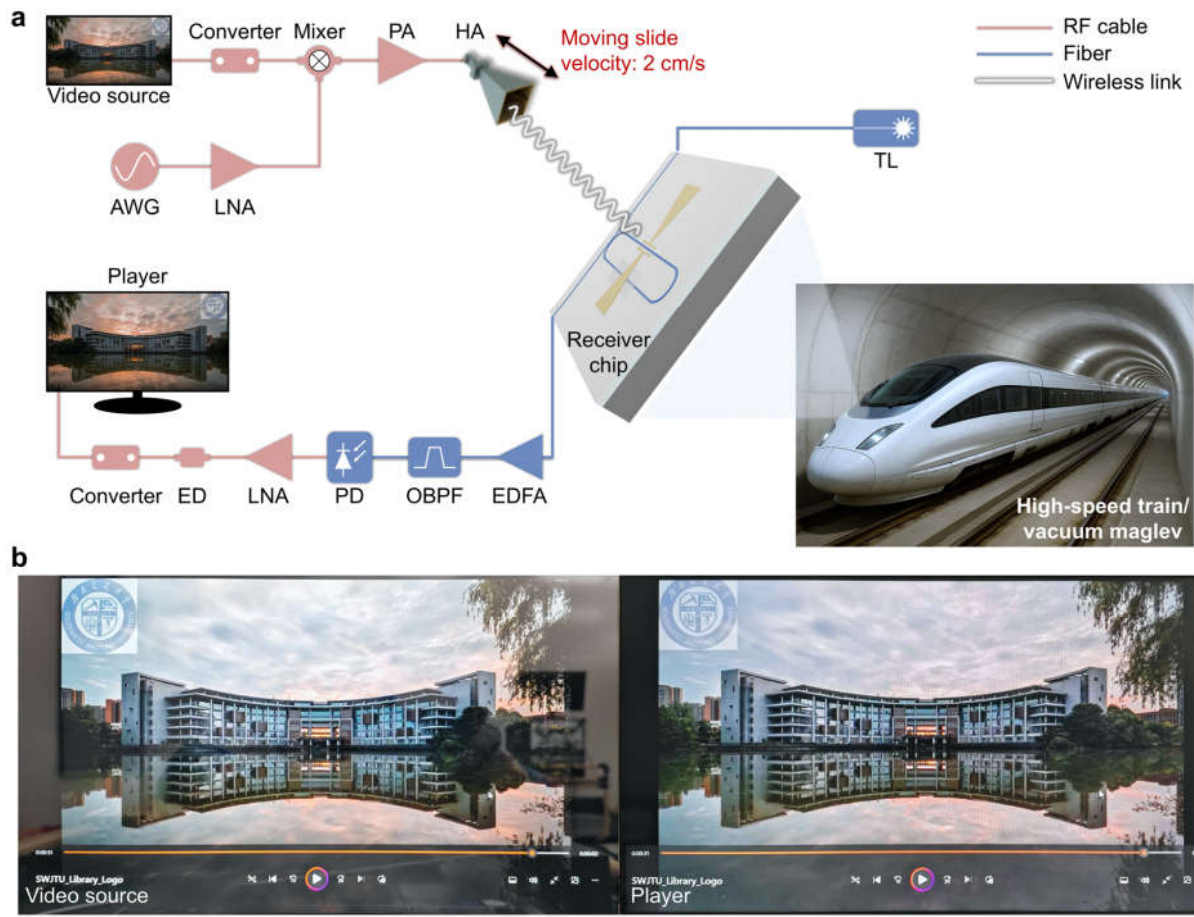

**Fig. 4. Demonstration of a real-time video transmission system for moving scenarios like high-speed trains or vacuum maglevs. a** Experimental setup of the real-time video transmission demonstration. **b** Photos showcasing the screenshot of HD video at the transmitter (left, static base station) and the player (right, moving vehicle-borne receiver). ED, envelope detector.

**Real-time video transmission in moving scenarios.** High-mobility scenarios represent another promising application for the synergistic photonic RF receiver, such as vehicle-borne RF frontends in high-speed trains and vacuum maglevs. The combination of high velocity and limited installation space brings crucial challenges for real-time RF communication. A real-time HD video streaming experiment is performed under moving condition (Fig. 4a). At the transmitter, an HD video source is upconverted to 30.5 GHz and then radiated into free space via a horn antenna. As the horn antenna moves forward or backward at a velocity of 2 cm/s, the photonic RF receiver continuously captures the radiated RF signal carrying the video stream. By demodulating the radiated RF signal encoded in the optical signal, the HD video source is recovered in real-time and synchronously displayed on another player (Fig. 4b). The distortion-free, real-time video transmission confirms the potential of the photonic RF receiver for future high-mobility scenario applications.

## 4. Conclusion

In this work, we demonstrate a monolithic photonic RF receiver with synergistic integration of a bow-tie antenna and a microring-enhanced modulator. Because of the dual-resonance enhancement mechanism, the monolithic and synergistic chip achieves a record-high FOM of 3.88 $W^{-1/2}$ within a compact footprint of only 2 × 1.7 $mm^2$, marking an impactful advancement toward both high conversion efficiency and miniaturized size for photonic RF receivers. Two representative experimental demonstrations, including the ISAC system and real-time HD video transmission, further highlight the chip's potential. This work unlocks the promising deployment of the photonic RF receiver in diverse emerging wireless communication and sensing applications, particularly in space-constrained, covert, and conformal environments.